\documentstyle[12pt]{article}
\begin{document}
\newtheorem{thm}{Theorem}
\newtheorem{cor}{Corollary}
\newtheorem{Def}{Definition}
\newtheorem{lem}{Lemma}
\begin{center}
%
{\large \bf Constants of the Motion in a Gravitational Field and the Hamilton-Jacobi Function} \vspace{5mm}

Paul O'Hara
\\
\vspace{5mm}
{\small\it
Dept. of Mathematics\\
Northeastern Illinois University\\
5500 North St. Louis Avenue\\
Chicago, Illinois 60625-4699.\\
\vspace{5mm}
email: pohara@neiu.edu}
\end{center}
\vspace{10mm}
\begin{abstract}

In most text books of mechanics, Newton's laws or Hamilton's equations of motion are first written down and then solved based on initial conditions to determine the constants of the motions and to describe the trajectories of the particles. In this essay, we take a different starting point. We begin with the metrics of general relativity and show how they can be used to construct by inspection constants of motion, which can then be used to write down the equations of the trajectories. This will be achieved by deriving a Hamiltonian-Jacobi function from the metric and showing that its existence requires all of the above mentioned properties. The article concludes with four applications, which includes a derivation of Kepler's First Law of Motion for planets, and a formula for describing the trajectories of galaxies moving in a space defined by the Robertson-Walker metric.

\end{abstract}
\vskip 10pt

Max Born in his book ``Natural Philosophy of Cause and Chance'' gives a derivation of Newton's laws of gravity from Kepler's laws of planetary motion noting that it ``is the basis on which [his] whole conception of causality in physics rests''(\cite{born},p.129).  In the spirit of that insight, the essay will explore the metrics of general relativity and show how it is possible to use them to derive both the constants of the motion and the particle trajectory in a gravitational field.

The key to this development will be rewriting the metric as an exact differential associated with the Hamiltonian-Jacobi function. Such exact differentials can always be constructed by noting that the inner product of any gradient vector $\nabla \psi(s)$ with a tangent vector to a curve $\textbf{ds}$ is always exact. More precisely, using spinor notation, a local tetrad can be constructed at any point on the curve (\cite{ohara}), with one-form ${\tilde ds}=\gamma^adx_a$
and its dual ${\tilde \partial_s \psi}=\gamma^a\frac{\partial \psi}{\partial x_a}$ such that
\begin{eqnarray}\frac{\tilde{ds}}{d\tau}.\tilde{\partial}_s\psi&=&\frac12\left\{\frac{\tilde{ds}}{d\tau},\tilde{\partial}_s\psi\right\}+\frac12
\left[\frac{\tilde{ds}}{d\tau},\tilde{\partial}_s\psi\right]\\
&=&\vec\frac{d \psi}{d\tau}+\frac{\textbf{ds}}{d\tau}\wedge\vec{\frac{\partial
\Psi}{\partial s}}.
\end{eqnarray}
Equations (1) and (2) can be identified by noting that the anti-commutator
and commutator relationships in $\frac{\tilde{ds}}{d\tau}.\tilde{\partial}_s\psi$ associated with the spinor tetrad, define a dot product and a cross product respectively.

We say that $\psi(\tau)\equiv W$ is a Hamilton-Jacobi function if (1) is true and $\left[\frac{\tilde{ds}}{d\tau},\tilde{\partial}_s W\right]=0$, or equivalently $W$ is a Hamilton Jacobi function whenever $dW=p^*_adx^a$ is an exact differential, where $p^*_a=\frac{\partial W(s)}{\partial x^a}=
W^{\prime}p_a$, and $p_a=\frac{\partial s}{\partial x}$. Usually, $p_o$ is denoted by $-H$ and $dW=p^*_1dx^1+p^*_2dx^2+p^*_3dx^3-H^*dt$. In particular, if $s$ is a parameter denoting length of a smooth curve then \cite{ohara}
\begin{equation} dW=p^*_1dx^1+p^*_2dx^2+p^*_3dx^3-H^*dt\ \ \textrm{iff}\ \ ds=p_1dx^1+p_2dx^2+p_3dx^3-Hdt.\end{equation}

It follows that Hamilton-Jacobi functions can be constructed at will starting from the metric.

Consider
\begin{equation} ds^2=g_{ij}dx^idx^j \end{equation}
this is equivalent to
\begin{equation} \frac{ds}{d\tau}ds=g_{ij}\frac{dx^i}{d\tau}dx^j.\end{equation}
The requirement that $dW\equiv \frac{ds}{d\tau}ds$ be a Hamilton-Jacobi function gives
\begin{equation} p_j=g_{ij}\frac{dx^i}{d\tau},\qquad  \tau\ \textrm{a parameter} \end{equation}
and that $\frac{\partial p_j}{\partial x^i}=\frac{\partial p_i}{\partial x^j}$ for all $i, j$.  Indeed, the simplest possible solutions occur when $p_i$ is independent of $j\neq i$ for all $j$, and $p_i=constant $ otherwise. Indeed, this can be written as a simple lemma:
\begin{lem} Let $ds^2=g_{ij}dx^idx^j$ and $p_i=g_{ij}\dot{x}^j=\frac{\partial s}{\partial x^i}$ be independent of $j$ for all $j\ne i$, if $p_j=p_j(x_i,\dot{x}_j)$ for some $j$ then $p_j=constant$ iff $\frac{\partial^2 s}{\partial x_j^2}=0.$
\end{lem}
{\bf Proof:} Since $s=s(x^i)$ then \begin{eqnarray*} \dot{p}_j&=&\frac{\partial p_j}{\partial x_i}\dot{x}_i\\
&=&\sum_{i\ne j}\frac{\partial^2 s}{\partial x_i \partial x_j}\dot{x}_i+\frac{\partial^2 s}{\partial x_j^2}\dot{x}_j.
\end{eqnarray*}
Also $p_i=g_{ij}\dot{x}^j=\frac{\partial s}{\partial x^i}$ is independent of $j$ for all $j\ne i$ implies $\frac{\partial^2 s}{\partial x_j\partial x_i}=0$.
But exact differentiability requires that $\frac{\partial p_j}{\partial x^i}=\frac{\partial p_i}{\partial x^j}$ for all $i, j$.
It now follows that $p_j=p_j(x_i,\dot{x}_j)$ is constant iff $\frac{\partial^2 s}{\partial x_j^2}=0.$ $\Box$

\begin{cor}If $\tau$ is proper time and $p_j=p_j(x_i,\dot{x}_j)=m_of(x_i)\dot{x}_j$ ($j\ne i$) is constant as in Lemma 1 above then $\dot{p}^j=0$ iff
$$\frac {d^2 x^j}{d\tau^2}+\frac{f^{\prime}(x^i)}{f(x^i)}\frac{d x^i}{d\tau}\frac{d x^j}{d\tau}=0,$$ which is an equation for geodesic motion with
$\Gamma^j_{ij}=\frac{f^{\prime}(x^i)}{f(x^i)}$.
\end{cor}
{\bf Proof:} Differentiate $p^j= constant$ and result follows from taking the derivative.$\Box$

In practice, as we shall see below, the imposition of this requirement on the metric will allow one to solve for those curves and determine those potentials for which momentum is conserved. It also allows us to read off the constants of the motion by inspection.

It should be noted that even if $p_i$ are not constant, the requirement that $s$ be a Hamilton-Jacobi curve in a given coordinate system allows one to determine all possible motions not involving spin or vortex motion. For example, the differential
\begin{equation} dW(s)=2\gamma(s)xydx+\gamma(s)x^2dy\end{equation}
is an exact differential for all parameterizations  $x=x(s)$, $y=y(s)$, where $\gamma(s)=\gamma(x^2y)$ is smooth. On the other hand, if we require that they are both constant they pick out a very specific family of curves (one for each $k$) associated with $x^2=2kxy$.
\vspace{5 mm}

\noindent{\bf Example 1:} As an application of the above theory, we begin by considering planar motion in Minkowski space with metric
\begin{equation} ds^2=dr^2 + r^2d{\theta}^2-c^2dt^2. \end{equation}
This can be written with respect to a parameter $\tau$ by
\begin{eqnarray} \frac{ds}{d\tau}ds = \frac{dr}{d\tau}dr + r^2 \frac{d\theta}{d\tau}d\theta-c^2\frac{dt}{d\tau}dt. \end{eqnarray}
The requirement that $s(\tau)$ be a Hamilton-Jacobi function means that there exist a class of integrable curves (with a $dot$ over the letters to indicate differentiation with respect to $\tau$) such that
\begin{equation} \frac{\partial \dot{r}}{\partial \theta}=
\frac{\partial r^2\dot{\theta}}{\partial r},\
\frac{\partial \dot{t}}{\partial \theta}=
\frac{\partial r^2\dot{\theta}}{\partial t},\ \  \frac{\partial \dot{t}}{\partial r}=
\frac{\partial \dot{r}}{\partial t}.
\end{equation}

By inspecting the metric, it is clear that $\frac{\partial s}{\partial r}$ and $\frac{\partial s}{\partial t}$ can be chosen independently of $\theta$. Consequently imposing the restriction $\frac{\partial^2 s}{\partial \theta^2}=0$ in accordance with the lemma (or equivalently choosing the action $s(r, \theta)=k\theta +f(r)$), exact differentiability requires that $r^2\dot{\theta}$ is a constant of the motion.
However, one could question whether this is the correct form for the equations of motion, especially in a non-Minkowski space. Returning to Equation (1), we see that exact differentiability requires that for the above metric $m_o\frac{\partial s}{\partial r}=m(s)\dot{r}$ and $m_o\frac{\partial s}{\partial \theta}=m(s)r^2\dot{\theta},$ where $m$ is a differentiable function of $s$, and can be identified with mass. It follows that geodesic motion associated with the metric $ds^2=dr^2 + r^2d{\theta}^2-c^2dt^2$ is given by $m(s)\dot{r}=p_r=constant$ and $m(s)r^2\dot{\theta}=constant$. Moreover, since any two constants are proportional to each other, an exact curve can be written down for this motion by noting that $m(s)\dot{r}=k_3m(s)r^2\dot{\theta}$, is equivalent to $-k_3r\theta+k_4r=1$.

On a final note, regarding this example, if we were to treat the problem from a classical perspective as a particle with rest mass $m_0$ moving in Minkowski space with radial acceleration defined by the covariant derivative $\frac{D\dot{r}}{ds}=\ddot{r}-r\dot{\theta}^2$, where $\dot{r}$ and $r^2\dot{\theta}$ are constant along a trajectory, then the acceleration would be given by
$a=-r\dot{\theta}^2=-(r^2\dot{\theta})^2/r^3=-k^2/r^3$. This defines an inverse cube and not an inverse square law of motion.
\vspace{5 mm}

\noindent{\bf Example 2:} As a second example consider the metric
\begin{equation}\dot{s}ds=\frac{l\dot{r}}{\sin\theta}dr+ r^2\dot{\theta}d\theta-c^2\dot{t}dt,\end{equation}
that can be derived from Kepler's first and second laws of planetary motion which states that planets move on ellipses given by $l/r=1+\epsilon \cos\theta$, with constant angular momentum. However, for the purpose of this essay, let us begin with the metric and require that $\dot{s}$ be an exact differential such that $p_r=\frac{\partial s}{\partial r}$ and $p_{\theta}=r^2\theta$ are constants of the motion, as in the case of geodesic motion. It immediately follows that $p_r=\epsilon p_{\theta}$ or equivalently that  $\frac{l\dot{r}}{\sin \theta}=\epsilon r^2\dot{\theta}.$ Integrating out gives the equation of a conic for an inverse square law of motion, which is Kepler's first law of motion.
\vspace{5 mm}

\noindent{\bf Example 3:} The same techniques can also be used to identifying the constants of the motion associated with all metrics in which the equations of motion obey the Hamilton-Jacobi equation. In the Schwartzschild space with a metric of the form
\begin{equation} ds^2=B(r)dt^2-A(r)dr^2-r^2d{\theta}^2-r^2\sin^2\theta d\phi^2, \end{equation} it is clear that all terms in the expansion except
the $A(r)dr^2$ contain mixed variables. Indeed, taking $\frac{\partial s}{\partial r}=A(r)\dot{r}$ to be independent of $t, \theta$ and $\phi$ gives
$$\frac{\partial A(r)\dot{r}}{\partial \theta}=\frac{\partial A(r)\dot{r}}{\partial \phi}=\frac{\partial A(r)\dot{r}}{\partial t}=0.$$
It now follows by Lemma 1 and the exact differentiability of the Hamilton-Jacobi function defined by
$s(r, \theta, \phi, t)\equiv k_1t+k_2+k_3\phi+f(r)$ that $p_t,p_{\theta}, p_{\phi}$ are constants of the motion such that $$B(r)\dot{t}=k_1, \theta=k_2, r^2\dot{\phi}=k_3.$$
These three constants are well known and can be easily shown to be associated with geodesic motion. Also by noting that any two constants can be related by a constant of proportionality $\epsilon$, if follows that any (geodesic) trajectory of the motion must obey the equation $k_1=\epsilon k_3$ or equivalently $\frac {B(r)}{r^2}=\epsilon \frac{d{\phi}}{dt}.$
\vspace{5 mm}

\noindent{\bf Example 4:} Similarly in the case of the Robertson-Walker metric
\begin{equation} ds^2=dt^2-R^2(t)\left\{\frac{dr^2}{1-kr^2}+r^2d\theta^2+r^2\sin^2\theta d\phi^2\right\}\end{equation}
there exists trajectories for which
$\dot{t}, R^2(t)\frac{\dot{r}}{1-kr^2}, R^2(t)r^2\dot{\phi}\ \textrm{and}\ \theta $
are constants of the motion, and in this case a generalize first law of Kepler would require that galaxies move on trajectories given by
$\frac{\dot{r}}{r^2(1-kr^2)}=\epsilon \dot{\phi}$. This can be integrated out, using partial fractions, to give the family of curves
\begin{equation} -\epsilon r \phi +k_5r+\frac{r\sqrt{k}}{2}\ln\left(\frac{1+\sqrt{k}r}{1-\sqrt{k}r}\right)=1.\end{equation}
\vspace{5 mm}

\noindent{\bf Conclusion:} There is something special about Hamilton-Jacobi functions. Not only can they be used to derive Hamilton's equations but they allow us to identify both equations and constants of motion in the spirit of Born's observation, and they also determine the trajectories in general for natural motions. In that regard, it should be recalled that if $s(\tau)$ is a Hamilton-Jacobi function then so also are smooth functions $W(s)$ and more sophisticated motions will require their use. For example in the case of simple harmonic motion associated with the metric of special relativity $ds^2=dx^2-c^2dt^2$, the requirement that $\dot{x}(s)=constant$ means that $W=A\cos(ks)$ will determine a simple harmonic motion.

 So what information is stored in a metric.  The answer is ``a lot.'' In the context of the overall field of mechanics the Hamiltonian-Jacobi functions  with gradient $\nabla W$ serve as gauge terms for the more general motion which can be written (see equation (2)) as
\begin{eqnarray}\frac{\tilde{ds}}{d\tau}.\tilde{\partial}_s\psi
=\frac{dW}{d\tau}+\vec A.\frac{d s}{d\tau}+\vec{A}\wedge\vec{\frac{ds}{d\tau}}.
\end{eqnarray}
But this is a discussion for another day.





\end{document}